\documentclass[review,number,sort&compress]{elsarticle}

\DeclareMathAlphabet{\mathpzc}{OT1}{pzc}{m}{it}
\usepackage{lineno}
\usepackage{setspace}
\usepackage{epstopdf}
\usepackage{graphicx}
\usepackage{amssymb}
\usepackage{amsmath}
\usepackage{amsthm}
\usepackage[colorlinks=true]{hyperref}
\usepackage{breakurl}
\usepackage[all]{hypcap}
\usepackage{mathrsfs}

\usepackage[percent]{overpic}
\usepackage{epsfig}
\usepackage{verbatim}
\usepackage{multirow}
\usepackage{float}
\usepackage{array}
\usepackage[T1]{fontenc}
\usepackage{url}
\usepackage{hyperref}
\usepackage{natbib}
\usepackage{wrapfig}

\journal{Nuclear Instruments and Methods A}
\begin{document}

\begin{frontmatter}
\title{Single channel PICOSEC Micromegas detector with improved time resolution}

\author[1]{A. Utrobicic \corref{cor1}}
\ead{antonija.utrobicic@irb.hr}
\author[10]{R. Aleksan}
\author[2]{Y. Angelis}
\author[3]{J. Bortfeldt}
\author[4]{F. Brunbauer}
\author[22,23]{M. Brunoldi}
\author[2]{E. Chatzianagnostou}
\author[5]{J. Datta}
\author[11]{K. Dehmelt}
\author[6]{G. Fanourakis}
\author[22,23]{D. Fiorina\fnref{fn1}}
\author[4,7]{K. J. Floethner}
\author[8]{M. Gallinaro}
\author[9]{F. Garcia}
\author[10]{I. Giomataris}
\author[11]{K. Gnanvo}
\author[10]{F.J. Iguaz\fnref{fn2}}
\author[4]{D. Janssens}
\author[10]{A. Kallitsopoulou}
\author[13]{M. Kovacic}
\author[11]{B. Kross}
\author[10]{P. Legou}
\author[4,14]{M. Lisowska}
\author[15]{J. Liu}
\author[7,16]{M. Lupberger}
\author[4,2]{I. Maniatis\fnref{fn3}} 
\author[11]{J. McKisson}
\author[15]{Y. Meng}
\author[4,16]{H. Muller}
\author[4]{E. Oliveri}
\author[4,17]{G. Orlandini}
\author[11]{A. Pandey}
\author[10]{T. Papaevangelou}
\author[18]{M. Pomorski}
\author[4]{L. Ropelewski}
\author[2,19]{D. Sampsonidis}
\author[4]{L. Scharenberg}
\author[4]{T. Schneider}
\author[10]{L. Sohl \fnref{fn4}}
\author[4]{M. van Stenis}
\author[20]{Y. Tsipolitis}
\author[2,19]{S.E. Tzamarias}
\author[22,23]{I. Vai}
\author[4]{R. Veenhof}
\author[22,23]{P. Vitulo}
\author[15]{X. Wang}
\author[4,22]{S. White}
\author[11]{W. Xi}
\author[15]{Z. Zhang}
\author[15]{and Y. Zhou}

\address[1]{Ruđer  Bošković Institute, Bijeni\v{c}ka cesta 54, 10000 Zagreb, Croatia}
\address[2]{Department of Physics, Aristotle University of Thessaloniki, University Campus, GR-54124, Thessaloniki, Greece}
\address[3]{Department for Medical Physics, Ludwig Maximilian University of Munich,  Am Coulombwall 1, 85748 Garching, Germany}
\address[4]{European Organization for Nuclear Research (CERN), CH-1211, Geneve 23, Switzerland}
\address[5]{Stony Brook University, Dept. of Physics and Astronomy, Stony Brook, NY 11794-3800, USA}
\address[6]{Institute of Nuclear and Particle Physics, NCSR Demokritos, GR-15341 Agia Paraskevi, Attiki, Greece}

\address[7]{Helmholtz-Institut für Strahlen- und Kernphysik, University of Bonn, Nußallee 14–16, 53115 Bonn, Germany}

\address[8]{Laboratório de Instrumentacão e Física Experimental de Partículas, Lisbon, Portugal}

\address[9]{Helsinki Institute of Physics, University of Helsinki, FI-00014 Helsinki, Finland}

\address[10]{IRFU, CEA, Université Paris-Saclay, F-91191 Gif-sur-Yvette, France}

\address[11]{Jefferson Lab, 12000 Jefferson Avenue, Newport News, VA 23606, USA}

\address[12]{LIDYL, CEA, CNRS, Universit Paris-Saclay, F-91191 Gif-sur-Yvette, France}

\address[13]{University of Zagreb, Faculty of Electrical Engineering and Computing, 10000 Zagreb, Croatia}

\address[14]{Université Paris-Saclay, F-91191 Gif-sur-Yvette, France}

\address[15]{State Key Laboratory of Particle Detection and Electronics, University of Science and Technology of China, Hefei 230026, China}

\address[16]{Physikalisches Institut, University of Bonn, Nußallee 12, 53115 Bonn, Germany}

\address[17]{Friedrich-Alexander-Universität Erlangen-Nürnberg, Schloßplatz 4, 91054 Erlangen, Germany}

\address[18]{CEA-LIST, Diamond Sensors Laboratory, CEA Saclay, F-91191 Gif-sur-Yvette, France}

\address[19]{Center for Interdisciplinary Research and Innovation (CIRI-AUTH), Thessaloniki 57001, Greece}

\address[20]{National Technical University of Athens, Athens, Greece}


\address[21]{University of Virginia, Virginia, U.S.A.}

\address[22]{Dipartimento di Fisica, Università di Pavia, Via Bassi 6, 27100 Pavia (IT)}
\address[23]{INFN Sezione di Pavia, Via Bassi 6, 27100 Pavia (IT)}

\cortext[cor1]{Correspondence to: Ruđer  Bošković Institute, Bijeni\v{c}ka cesta 54, 10000, Zagreb, Croatia.}
\fntext[fn1]{Now at Gran Sasso Science Institute, Viale F. Crispi, 7 67100 L'Aquila, Italy}
\fntext[fn2]{Now at SOLEIL Synchrotron, L’Orme des Merisiers, Départementale 128, 91190 Saint-Aubin, France.}
\fntext[fn3]{Now at Department of Particle Physics and Astronomy, Weizmann Institute of Science, Rehovot, 7610001, Israel.}
\fntext[fn4]{Now at TÜV NORD EnSys GmbH Co. KG.}

\begin{abstract}
This paper presents design guidelines and experimental verification of a single-channel PICOSEC Micromegas (MM) detector with an improved time resolution. The design encompasses the detector board, vessel, auxiliary mechanical parts, and electrical connectivity for high voltage (HV) and signals, focusing on improving stability, reducing noise, and ensuring signal integrity to optimize timing performance. A notable feature is the simple and fast reassembly procedure, facilitating quick replacement of detector internal components that allows for an efficient measurement strategy involving different detector components. The paper also examines the influence of parasitics on the output signal integrity. To validate the design, a prototype assembly and three interchangeable detector boards with varying readout pad diameters were manufactured. The detectors were initially tested in the laboratory environment. Finally, the timing performance of detectors with different pad sizes was verified using a Minimum Ionizing Particle (MIP) beam test. Notably, a record time resolution for a PICOSEC Micromegas detector technology with a CsI photocathode of 12.5$\pm$0.8~ps was achieved with a 10~mm diameter readout pad size detector.
\end{abstract}
\end{frontmatter}

\newpage

\section{Introduction}
\label{sec:chap1}

Modern particle physics experiments have increased demand for high precision time, energy and position measurements that resulted in extensive detector research and development. Requirements, such as precise timing at the level of $\mathcal{O}$(10~ps) can be beneficial for pile-up rejection and improved particle identification at future accelerators \cite{ecfa, white2013experimental}.

\begin{figure}[!htb]
\begin{center}

\includegraphics[width=0.52\columnwidth]{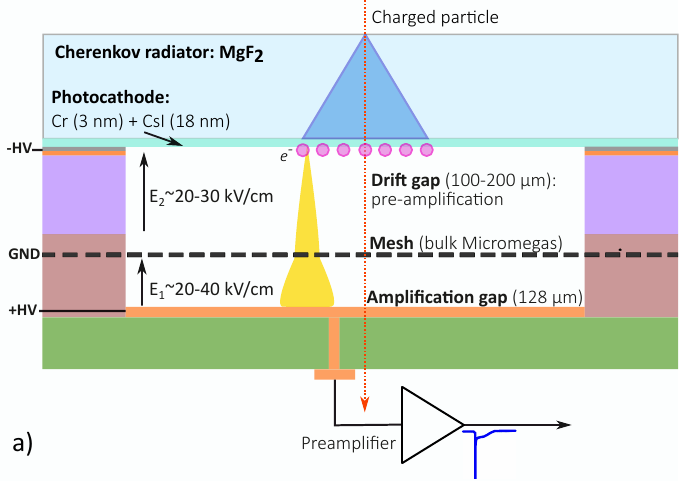}
\hspace{0.01cm}
\includegraphics[width=0.46\columnwidth]{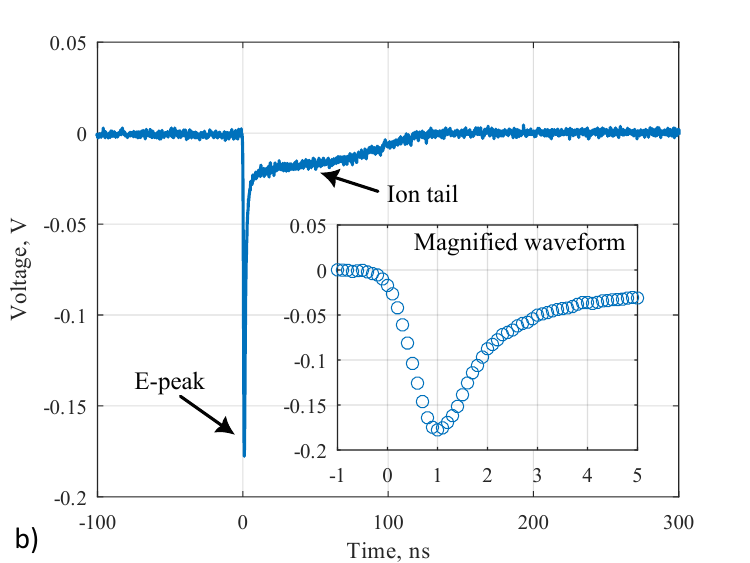}

\caption{a) Operating principle of the PICOSEC MM detector. b) Typical PICOSEC MM signal after amplifier.}
\label{fig_concept}
\end{center}
\end{figure}

PICOSEC Micromegas (MM) is a precise timing gaseous detector with a two-stage Micromegas amplification structure and semi-transparent photocathode coupled to a Cherenkov radiator \cite{papaevangelou2018fast,bortfeldt2018picosec}. The operation principle is based on overcoming the time jitter from the standard operation mode of MM detectors that occurs due to the different distance of ionization clusters in the drift region above the mesh \cite{giomataris1996micromegas}. The introduction of the Cherenkov radiator and photocathode in front of the gas volume with MM amplification structure enabled approximately synchronous ejection of all primary electrons at equal distances from a mesh. The longitudinal diffusion and probability of ionization by the passing particle were minimized by the reduction of drift gap thickness to a level of 100-200~$\mu$m \cite{bortfeldt2018picosec, derre2000fast}. The cross-section of the detector with the sketch of the operational principle is shown in Figure \ref{fig_concept}. The passage of a charged relativistic particle through the detector creates UV light in the Cerenkov radiator which is converted into primary electrons by a photocathode. Primary electrons are initially pre-amplified in the drift region of 200~$\mu$m length and finally amplified in the amplification region with the length of 128~$\mu$m, see Figure \ref{fig_concept} (a). Movement of the amplified charge induces a signal on an anode board that is composed of the electron peak with fast (sub ns) rise time component and slow ion tail (100 ns) component, see Figure \ref{fig_concept} (b).

The first PICOSEC MM detector was a ($\varnothing$1~cm) single channel prototype operating in a gas mixture of Ne:C$_2$H$_6$:CF$_4$ (80:10:10) at nominal temperature and pressure. It consisted of a 3~mm thick MgF$_2$ window as Cherenkov radiator with an 18 nm CsI semitransparent photocathode deposited on top of the 5.5 nm thick conductive Chromium layer. The bulk MM amplification structure was coupled to the MgF$_2$ window facing toward the photocathode. It consisted of a stainless steel calendered mesh, with 18 $\mu$m diameter wires and 45 $\mu$m openings, that was positioned in between the photocathode (at 200 $\mu$m distance) and the readout pad (at 128 $\mu$m distance) thus defining drift and amplification gap thicknesses. The prototype in this detector configuration achieved the best time resolution of 24~ps for measurements with 150~$\text{GeV}/c$ muons and a single photoelectron time resolution of 75~ps \citep{bortfeldt2018picosec}. Reduction of a drift gap thickness from 200~$\mu$m to 100~$\mu$m improved single photoelectron time resolution to 50~ps \cite{sohl2020single, bortfeldt2021modeling}. In the last few years the intensive detector R\&D continued towards covering larger areas, improving detector stability, robustness, and timing performance \cite{aune2021timing, kordas2020progress,manthos2020recent, sohl2019picosec}. 

Progress to a larger area coverage yielded new challenges in detector development. To obtain uniform time response over larger areas it was necessary to ensure uniform drift gap thickness within 10~$\mu$m over an entire active area. Besides timing performance, uniformity of the drift gap is crucial for the detector's stable operation. In this case, possible sources of drift gap non-uniformity, such as bending of the MM board or MgF$_2$ crystal needed to be addressed during the design phase. Recent developments included a 100-channel detector prototype with 100 cm$^2$ active area based on a hybrid ceramic MM board and with minimal mechanical coupling of all the active detector components to the housing \cite{utrobicic2023large}. The prototype was commissioned with minimum ionizing particles at the SPS H4 beamline and preserved the timing properties of a single-channel device \cite{maniatis2022research}. Operating with a thinner drift gap thickness of 180~$\mu$m and together with a custom-developed RF amplifier yielded a time resolution of 17~ps in the central pad region. The time resolution of 17-18~ps was obtained over almost the entire active area and 20.9$\pm$0.5~ps, when the signal is shared within $\pm$~1mm from the cross of the four adjacent pads, with Artificial Neural Network method analysis \cite{chatzianagnostou2022study}. Promising timing characteristics motivate future detector research regarding full detector readout chain, robustness, stability, and timing performance \cite{kallitsopoulou2021development,lisowska2023towards,janssens2024resistive}. 

  The primary objective of the improved single-channel PICOSEC MM detector is to ensure that its timing performance is mainly determined by the underlying physics, rather than limitations originating from surrounding components. This approach allows for testing of detector concept’s inherent timing limitations. Moreover, a small, easy-to-handle, high voltage (HV) stable detector prototype with the possibility for a simple and quick exchange of main detector components and good signal integrity would simplify the detector research and development.

\section{Considerations for single channel PICOSEC MM detector design}

\label{sec:chap2}
\subsection{HV stability}
Stable operation at high voltage is an important factor in the design of every gaseous detector. To fully exploit the detector and test the operational limits, one of the design objectives was to avoid discharge formation that might originate outside the detector's active region. By doing this, the source of instability can only come from the detector's operational parameters, and the charge developed in the active area avoiding external effects. 
The necessity for the frequent replacement of the internal parts and quick reassembly of the detector requires to use electrical connection techniques that will facilitate the process. In first prototypes of single-channel detectors, the electrical connections to external equipment were made by soldering. This proved critical in terms of required time, cleaning process and formation of sharp edges that could degrade HV stability. Moreover, the process itself was prone to human errors, often causing problems with running the detector after assembly. 

\begin{figure}[!htb]
\begin{center}
\includegraphics[width=0.59\columnwidth]{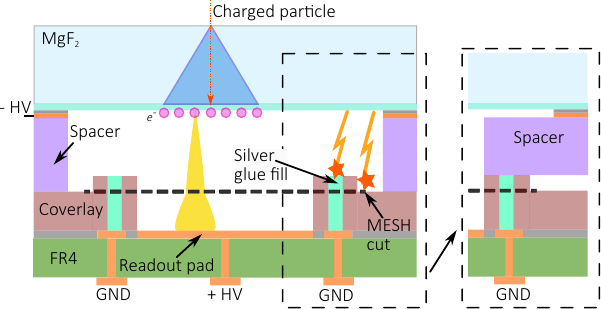}
\caption{ Detector cross-section with details on electrical connection points. Regions (indicated as silver glue fill and mesh cut) with higher electric fields prone to discharge formation are also shown. An example of potential design improvement by covering exposed metallic regions with the spacer's dielectric material (purple) is shown in the boxed panel on the right side.}
\label{fig_stability}
\end{center}
\end{figure}

Besides standard HV stability considerations, such as the clearance between the electrodes and housing, a detector layout specific for PICOSEC MM had to be considered in detail. Although the reduction of a drift gap thickness to a level of 100~$\mu$m is beneficial for the timing properties of the detector, it poses a large risk for electrical discharges in both the active area and the outside in the region between the coverlay and the photocathode (see Figure \ref{fig_stability}). The two most critical locations for the photocathode discharge formation are the electrical connection of mesh to the ground and mesh termination area. Bulk MM \cite{giomataris2006micromegas} are produced by embedding a mesh between two thin insulating layers, and the mesh is electrically connected to the ground copper pad using silver glue. The mesh is usually terminated at the perimeter, a few mm away from the active area, by mechanical cutting it, leaving thin sharp wire ends which can be exposed to the photocathode.  This can result in local instabilities, as indicated in Figure \ref{fig_stability} (left). The design choice for the new detector was to reallocate these mesh connection points and mesh cut perimeter outside the photocathode area and/or to cover them with a dielectric barrier. To avoid adding additional parts, an already existing polyimide copper-clad spacer placed on top of the MM board to define a drift gap thickness and ensure electric connection (Figure \ref{fig_stability}, right), can be designed to also function as a barrier, thereby improving stability in the external region.

\subsection{Signal integrity}
\label{sec:sig_int}
Contribution to a time resolution comes from different factors that can be expressed as 

\begin{equation}
	\sigma^2=\left[\dfrac{V_t­}{dV/dt}\right]^2_{RMS} + \left[\dfrac{N}{dV/dt}\right]^2 +  \sigma^2_{arrival} + \sigma^2_{distortion} + \sigma^2_{digitization}.
\label{eq_time_resolution}
\end{equation}

The first term is called time-walk contribution and comes due to the timing dependence on signal amplitude, where larger signals will intersect the fixed threshold level earlier than the smaller ones. The most often used method for compensation of the time-walk is the Constant Fraction Discrimination (CFD) technique which determines the time-stamp at the fixed fraction of the signal amplitude. The second term denotes a time jitter caused by the electronic noise superimposed on the recorded signal, while the third term is related to the variance of signal arrival time originating from the non-uniformity of the preamplification gap and the difference in the distance of the primary photoelectron has to pass before creating an avalanche in the pre-amplification region \cite{bortfeldt2021modeling, aune2021timing,cartiglia2014performance, janssens2024resistive}. One approach to minimize the contribution of $\sigma_{arrival}$ is to reduce the drift gap thickness which enables the application of larger drift fields and reduction of the primary photoelectron drift length \cite{bortfeldt2021modeling}. However, reducing the drift gap thickness will cause mechanical non-uniformities to be more pronounced and thus require drift gap precision at a level of few $\mu$m \cite{utrobicic2023large}. The last two terms are related to the non-uniformity of the electrical field over the active area and to signal digitization. To improve or maintain good timing properties, each of these contributions needs to be considered in the detector and readout chain design. Signal characteristics such as low noise, large signal amplitude, and fast sub ns rise time are crucial to achieve this goal. 
  
One of the measures to improve signal quality and detector time resolution is the reduction of the electronic pick-up noise coming from an external RF background or high-voltage power supply. In a 100-channel detector prototype, a multi-layer printed circuit (PCB) board, the so-called Outer Board (OB), was used for the detector chamber sealing and routing of the high voltage and signal lines \cite{utrobicic2023large}. The ground planes poured in the multiple layers of the OB and their good electrical connection to the aluminium housing additionally improve immunity to the external RF noise. 

Moreover, noise originating from the HV power supply can be reduced with the use of a low-pass filter that is closely coupled to the detector HV connection points. These filters are integrated into the Outer PCB to improve immunity and simplify detector operation.

\begin{figure}[!htb]
\begin{center}
\includegraphics[width=0.69\columnwidth]{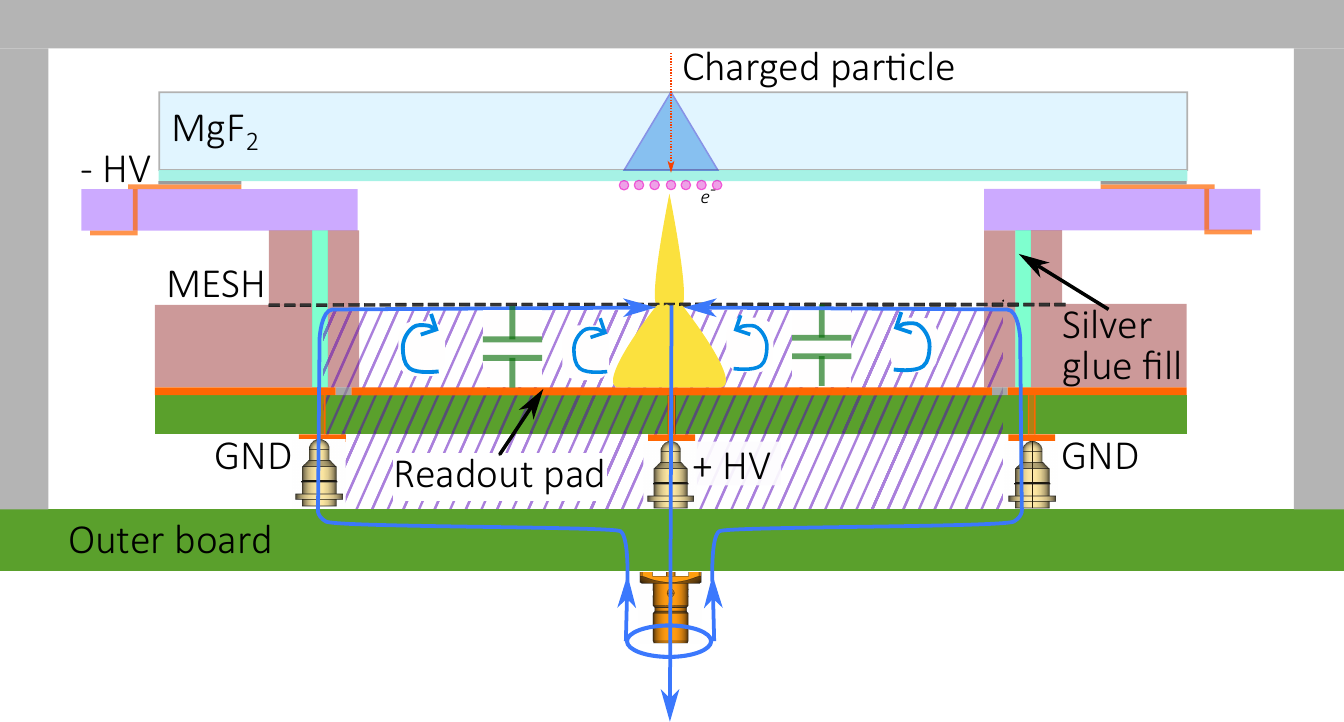}
\caption{Sketch of the detector cross-section including capacitance and inductance elements due to the detector geometry. }
\label{fig_cross_L_C}
\end{center}
\end{figure}

As indicated by equation \ref{eq_time_resolution}, a high slew rate of the signal at the discriminator plays an important role in achieving good timing properties. The slew rate depends on the dynamics of the induced current at the readout pad during the electron peak formation, the bandwidth of the amplifier, and digitizer and the energy stored in parasitic components of the detector assembly, see Figure \ref{fig_cross_L_C}. The shape of the induced current is defined by the dynamics of the avalanche formation in a thin drift (pre-amplification) region and its final amplification at the mesh \cite{bortfeldt2021modeling}. Consequently, the increase in the dynamics of the induced signal can be influenced by using larger drift fields and smaller drift lengths.

The fast current signal is shunted by the detector capacitance, thus immediately reducing the dynamic properties of the signal at the output. Detector capacitance is dominated by the capacitance between the readout pad and the mesh, due to the relatively large surface and thin gap between those two electrodes, see Figure \ref{fig_cross_L_C}. Another contribution to the slew rate, which is often overlooked, is the parasitic inductance of the signal path. This originates from the loop encompassed by the signal output and return current paths, marked by the hatched area in Figure \ref{fig_cross_L_C}. The existence of parasitic inductance can negatively influence the slew rate and cause ringing in the output signal.

\begin{figure}[!htb]
\begin{center}
\includegraphics[width=0.99\columnwidth]{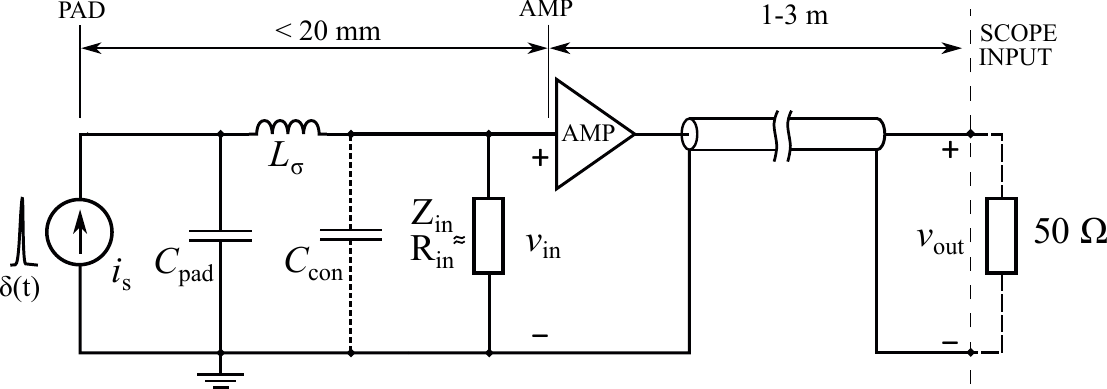}\\
\label{fig_schema_L_C}
\caption{Simplified circuit diagram of the detector and signal path}
\end{center}
\end{figure}

To explore this, a lumped element circuit has been derived to analyze the dependence of the signal response of the detector on the parasitic components, see Figure \ref{fig_schema_L_C}. The induced signal is modelled as a current source $i_s$  parallel to the pad capacitance $C_{pad}$. Stray inductance is modelled as a single inductor $L_{\sigma}$ placed between the readout pad and the signal output coaxial connector. The transmission line effects were neglected as the expected bandwidth of the detector is below 1~GHz, and the distances between the readout pad and the amplifier are lower than $\lambda/10$.

To simplify the analysis, and exclude the influence of the detector physics on the signal dynamics, a current source $i_s$ is modelled as the unit impulse. This can be justified if the induced electron impulse is very fast, which is usually true for an MM operating in PICOSEC mode where a high electric field is applied in the drift ($\approx$20-40~kV/cm) and amplification ($\approx$20-30~kV/cm) gap. A time response of the voltage at the input of the amplifier to a current impulse with a total charge of $Q_{pad}$ can be derived if the amplifier input impedance is taken to contain only resistance $R_{in}$:

\begin{equation}
\begin{split}
v_{in}\left(t\right)=&Q_{pad}\cdot\dfrac{R_{in}}{L_{\sigma}C_{pad}}\cdot\dfrac{1}{\sqrt{\dfrac{1}{L_{\sigma}C_{pad}}-\dfrac{R_{in}^2}{4L_{\sigma}^2}}}\cdot e^{-\dfrac{R_{in}}{2L_{\sigma}}\cdot t}\cdot \\ 
\cdot&\text{sin} \left(\sqrt{\dfrac{1}{L_{\sigma}C_{pad}}-\dfrac{R_{in}^2}{4L_{\sigma}^2}} \cdot t\right).
\end{split}
\label{eq_impulse_response}
\end{equation}

The maximum slew rate of the signal at the input of the amplifier in case of the impulse response occurs at $t = 0$~ns and its value can be expressed as:

\begin{equation}
\left|\dfrac{dv_{in}}{dt}\right|_{max}=Q_{pad}\cdot\dfrac{R_{in}}{L_{\sigma}C_{pad}}.
\label{eq_max_rise}
\end{equation}

\begin{figure}[!htb]
\begin{center}
\includegraphics[width=0.49\columnwidth]{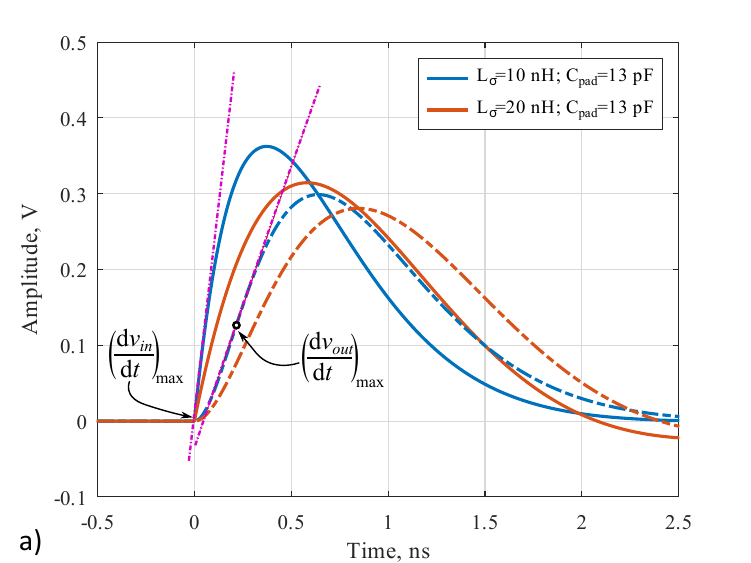}
\includegraphics[width=0.49\columnwidth]{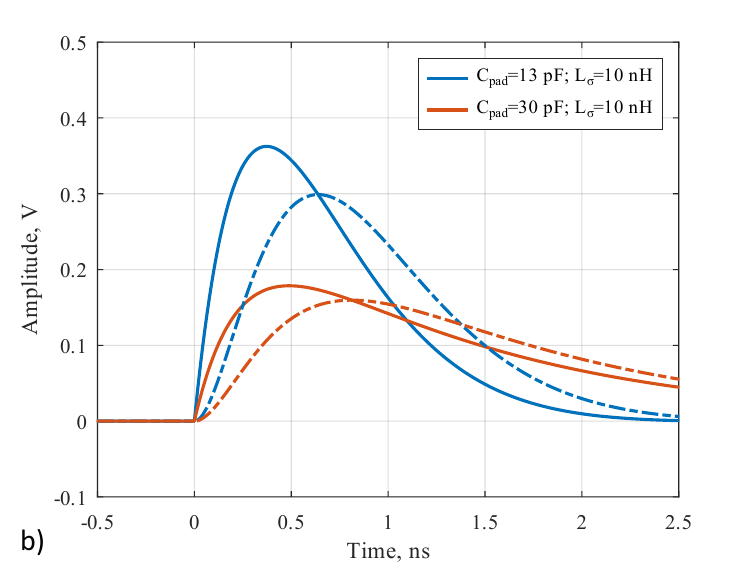}

\caption{a) Current impulse response for a fixed detector capacitance of 13~pF for inductance values of 10 and 20~nH. b) Current impulse response for fixed stray inductance of 10~nH for pad capacitance values of 13 and 30~pF. Solid lines represent the waveforms of the voltage at the input of the amplifier while dashed are calculated assuming the ideal 650~MHz bandwidth of the amplifier with unity gain. In this example a current impulse with total charge of $Q_{pad}=6.6$~pC was used as an excitation.}
\label{fig_impulse_response}
\end{center}
\end{figure}

It can be concluded that the slew rate depends on both the pad capacitance and pad inductance. Figure \ref{fig_impulse_response} shows the waveforms obtained using expression \ref{eq_impulse_response}. 
In addition to its influence on the slew rate, the pad capacitance also negatively affects the signal amplitude and, by extension, the signal-to-noise ratio (SNR). This highlights the importance of minimizing pad capacitance.
Moreover, further analysis shows that the negative effects of increased pad capacitance and stray inductance can be observed at the output of the idealized unity gain 650~MHz amplifier.

\section{Design of single-channel PICOSEC MM with metal anode}
\label{sec:chap3}
\subsection{Design of the Micromegas board and spacer}

The design of a new small channel detector has been driven by preserving or improving signal integrity, HV stability, and uniformity of time response. 

\begin{figure}[!htb]
\begin{center}
\includegraphics[width=0.49\columnwidth]{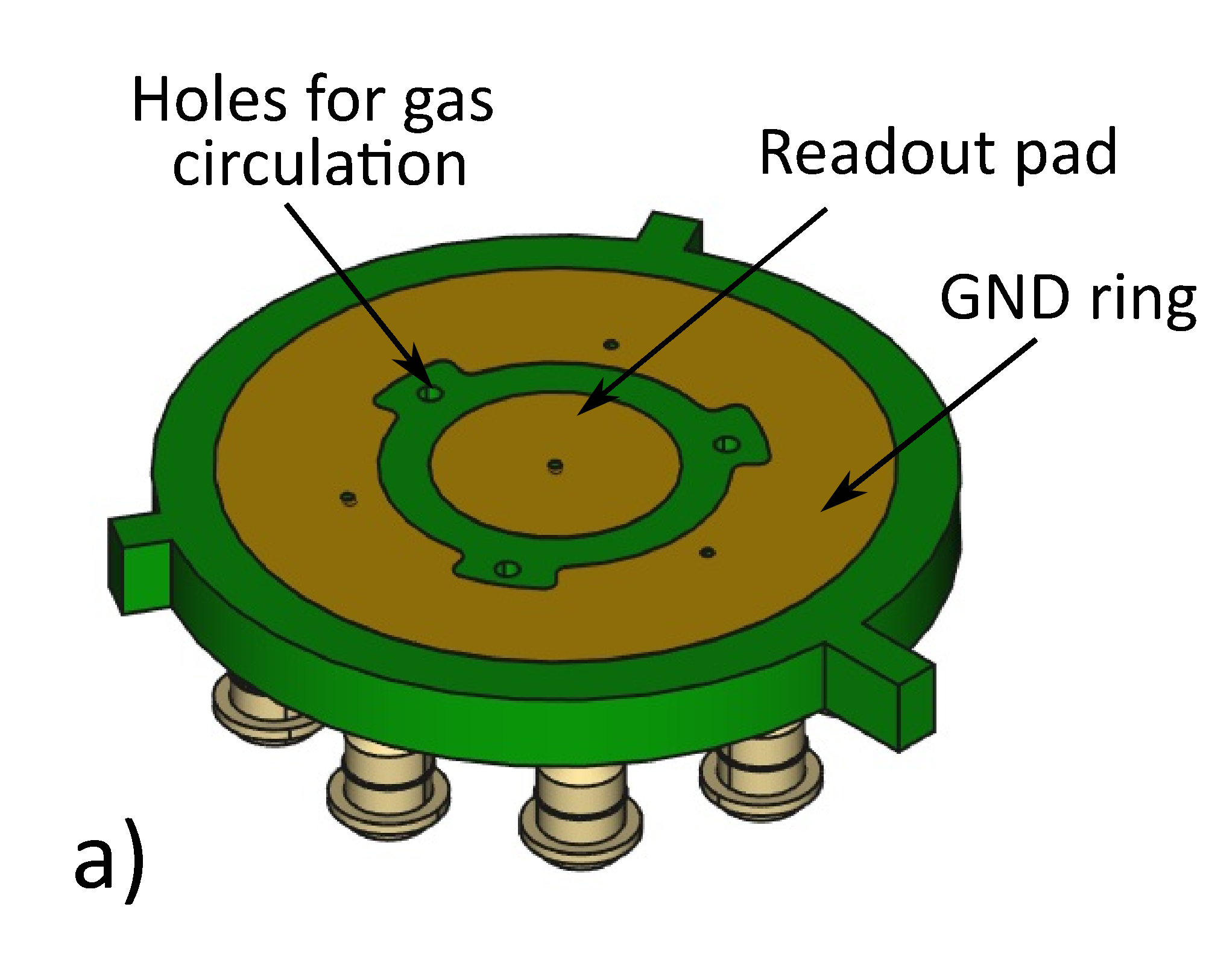}
\includegraphics[width=0.49\columnwidth]{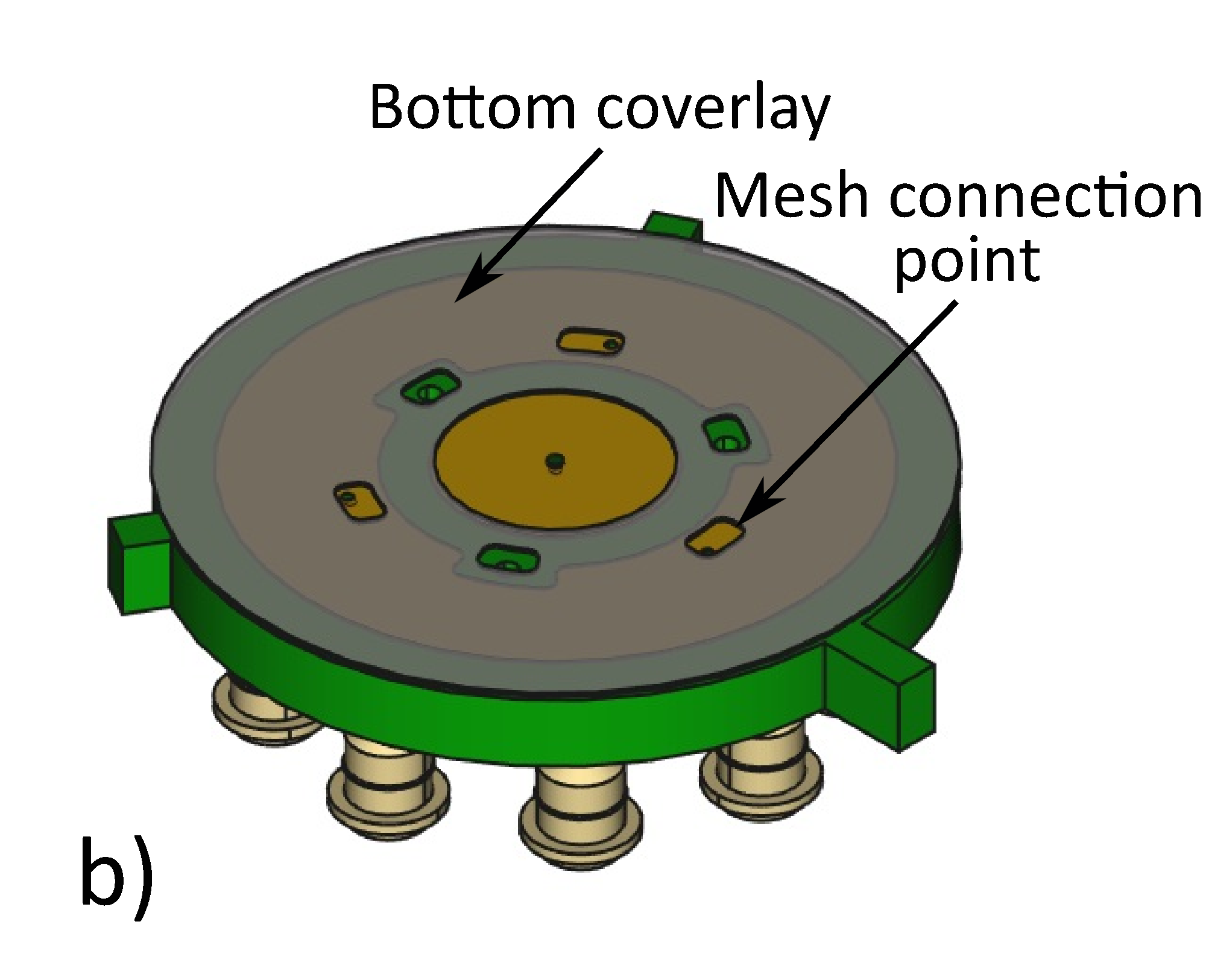}

\includegraphics[width=0.49\columnwidth]{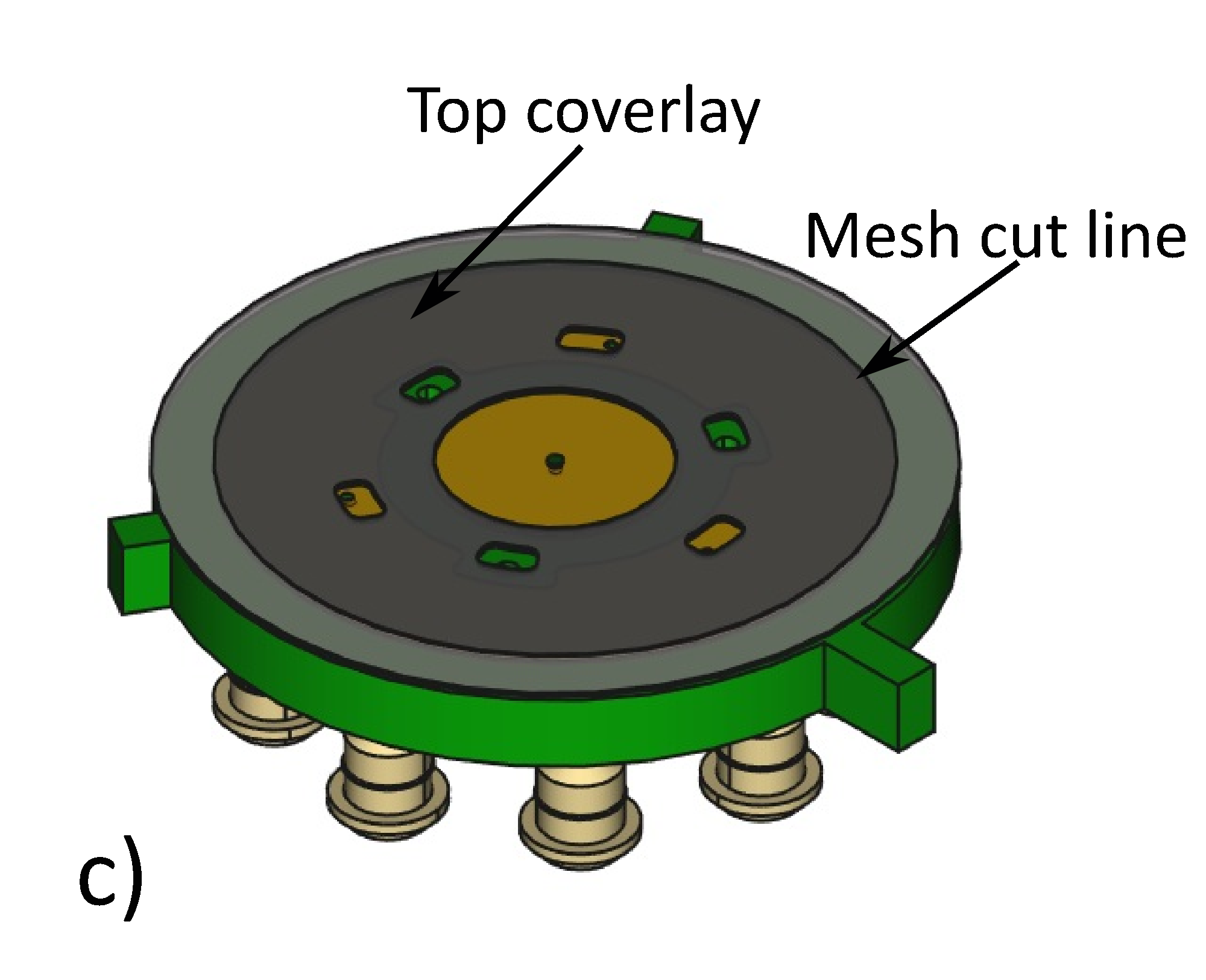}
\caption{3D model of the Micromegas board top side layout. a) Top copper layer that includes a readout pad in the middle, surrounded by a GND plane ring with three holes between them for gas circulation. b) First insulating layer where a circular opening exposes the readout pad,  along with several smaller openings for the mesh connection points to GND and for gas circulation. c) Second insulation layer placed on top of the Micromegas mesh. Its perimeter defines the mesh cut positions.}
\label{fig_det_top}
\end{center}
\end{figure}

A MM board was produced using a 3.2 mm thick FR4 printed circuit board with two copper layers. A copper readout pad ($\varnothing$10, $\varnothing$13 or $\varnothing$15~mm) and a surrounding ground (GND) plane ring are placed on the top side of the board, see Figure \ref{fig_det_top} (a). The bottom side of the board features the pads at spring-loaded pin landing positions which are connected to the top side readout pad and GND ring using filled and capped vias. One signal (central) spring-loaded pin and 9 GND connection spring-loaded pins are used in the coaxial arrangement to minimize stray inductance of the signal path.

The stainless steel mesh is integrated on top of the board in between two polyimide covering layers (coverlays). The bottom coverlay has a thickness of 127~$\mu$m and 32~mm outer diameter. It features the openings for the active area, gas circulation, and connection of the mesh to the GND ring, as shown in Figure \ref{fig_det_top} (b). The 51~$\mu$m thick top coverlay, laminated over the mesh, has the same openings as the bottom coverlay, but a smaller outer diameter of 27~mm. The mesh extends to the edge of the top coverlay, where a mesh cut is performed (Figure \ref{fig_det_top}. (c)).

\begin{figure}[!htb]
\begin{center}
\includegraphics[width=0.39\columnwidth]{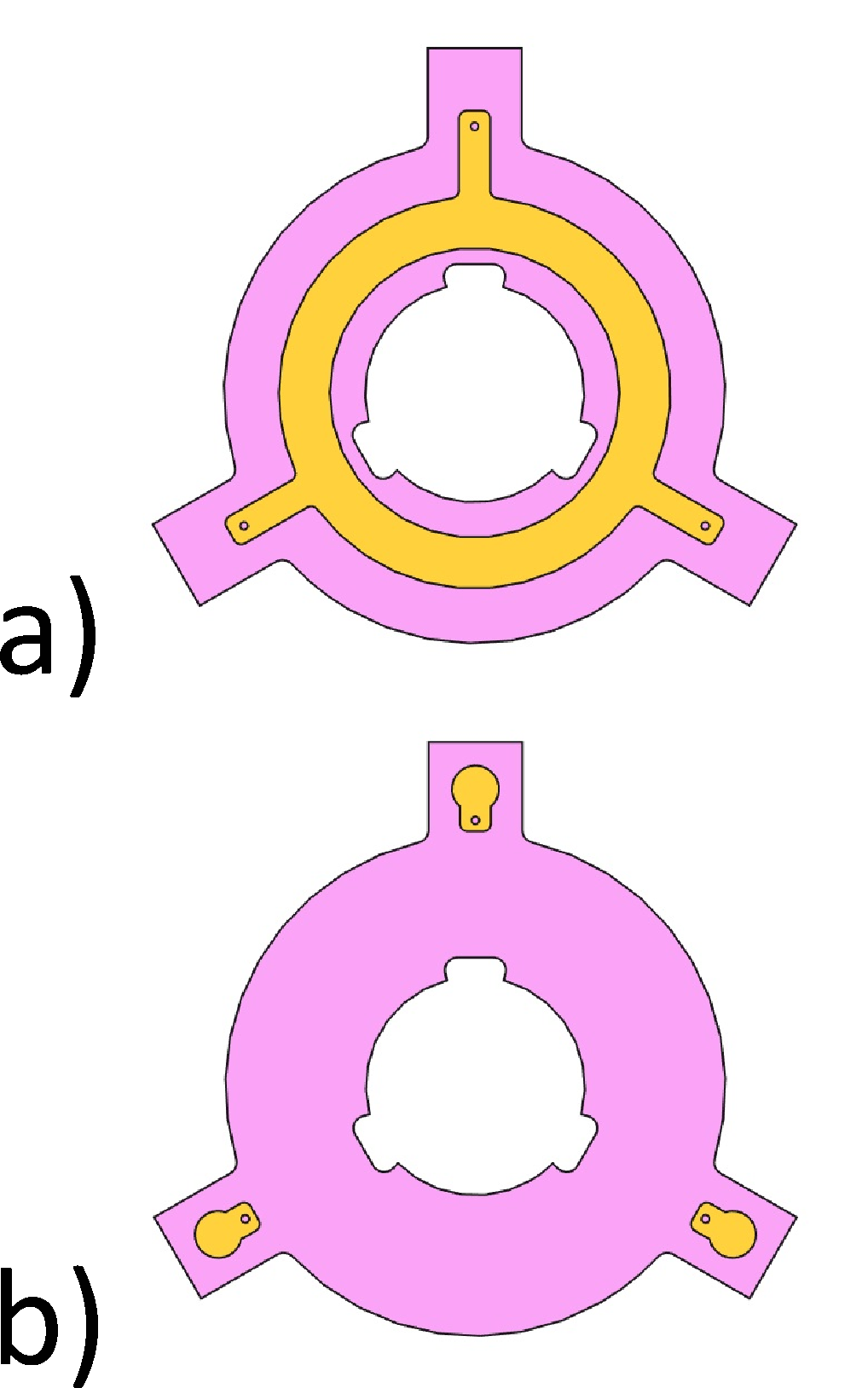}
\includegraphics[width=0.6\columnwidth]{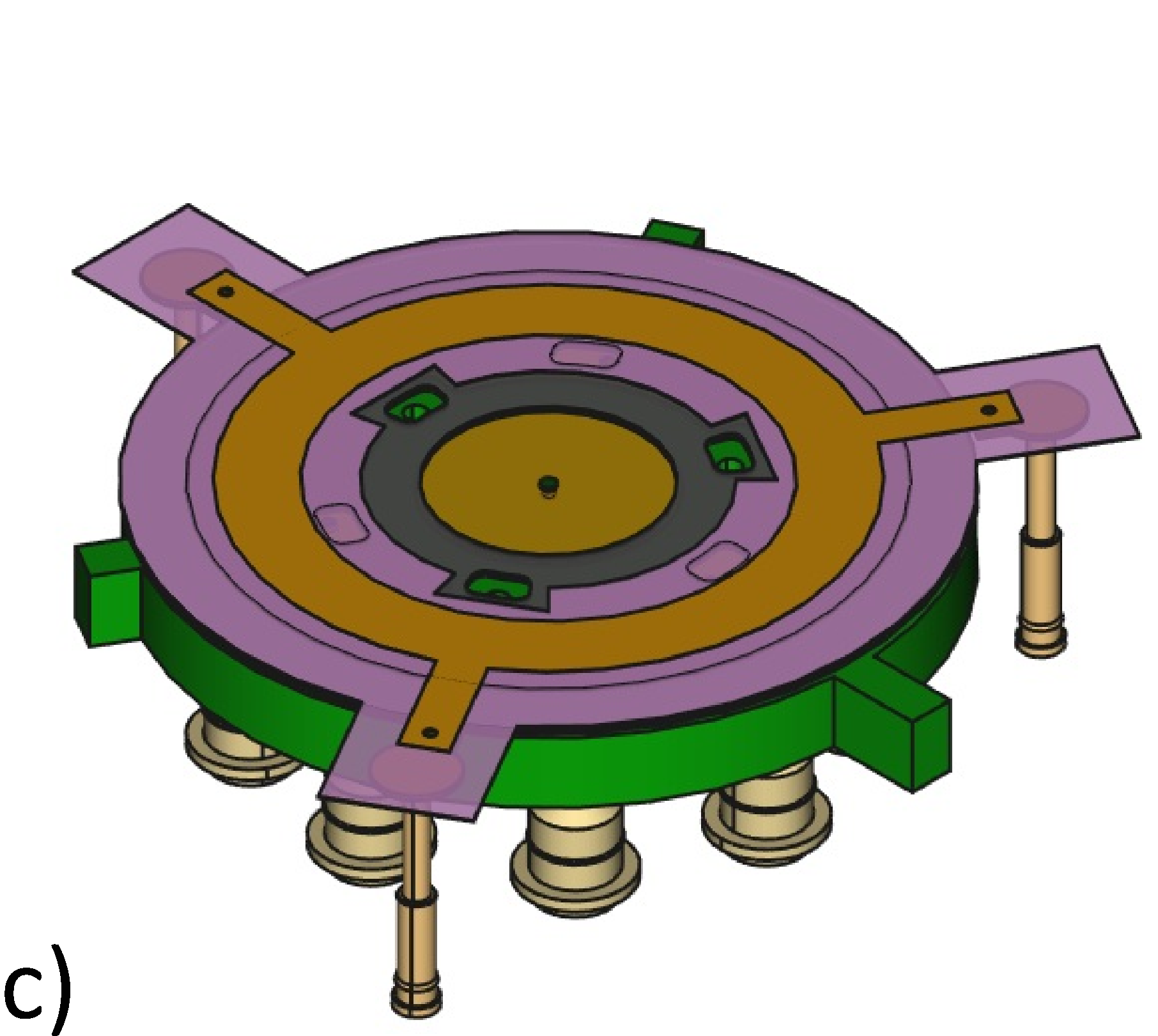}

\includegraphics[width=0.6\columnwidth]{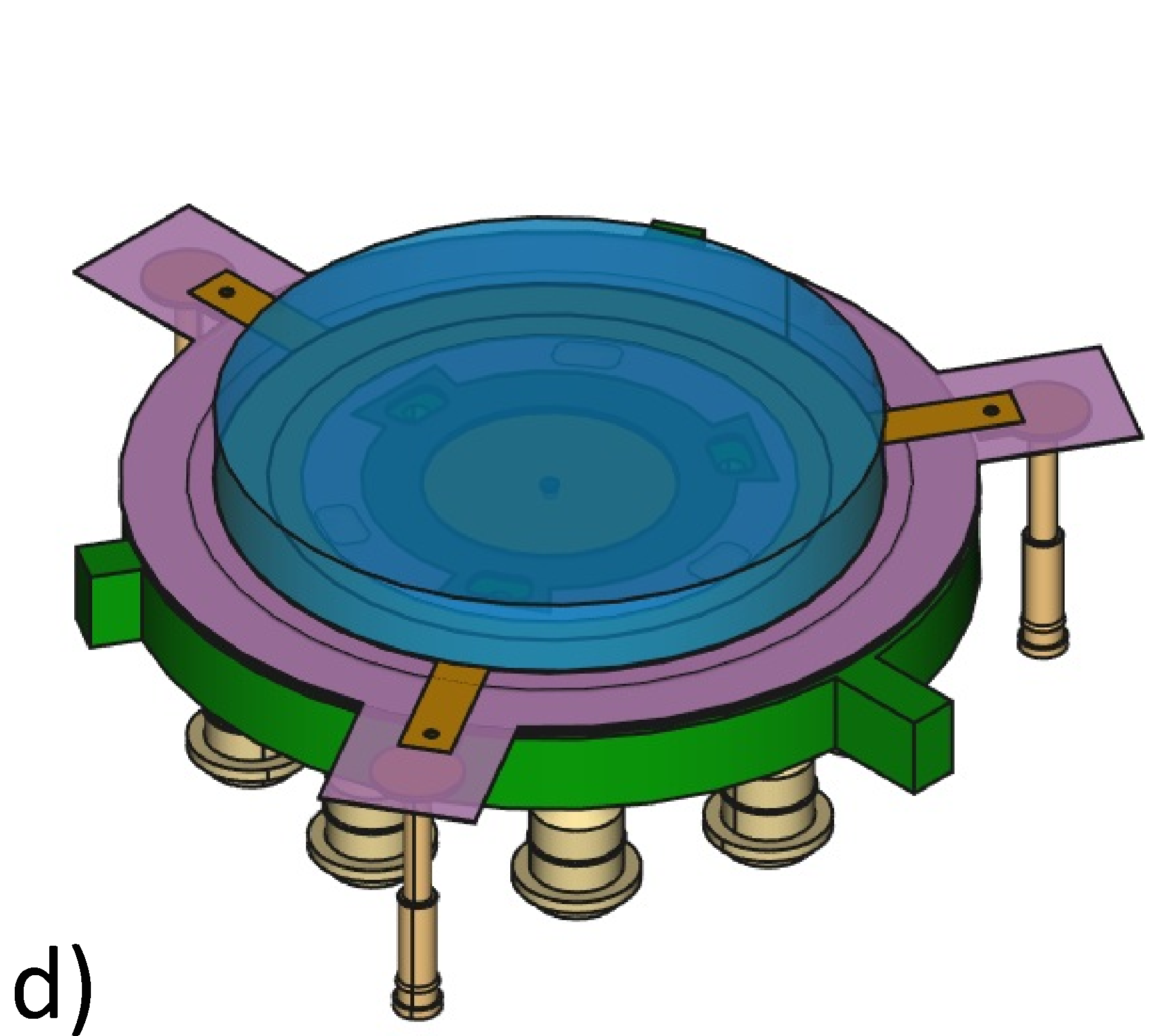}
\caption{Sketch of Micromegas board, spacer and MgF$_2$ crystal arrangement within the detector. a) Spacer top side layout. b) Spacer bottom side layout. c) Sketch of spacer mounting on top of the Micromegas board. d) Sketch of MgF$_2$ crystal placement on top of the spacer.}
\label{fig_det_spacer}
\end{center}
\end{figure}

Besides the crucial role, that ensures the stability of the detector, together with the top insulating layer, a spacer is used to mechanically define the size of the drift gap and establish the electrical connection of the cathode to the HV. Installation of the spacer has to be simple and fast without the need for any soldering. The spacer is made from a 70~$\mu$m thin two-layer copper-clad polyimide. The top side layer contains a copper ring with three extensions, connected to the small pads on the spacer bottom side, see Figure \ref{fig_det_spacer} (a) and (b). It is placed on the top coverlay and extends over the mesh edge and connection points to GND as shown in Figure \ref{fig_det_spacer} (c). In this way, the probability of a discharge formation from sharp wires at the mesh cut line or tips of conductive epoxy at the openings for mesh connection to the GND is reduced.
A MgF$_2$ crystal with a photocathode and conductive ring is placed on the spacer's top side to establish the electrical connection of the photocathode to the HV through three spring-loaded pins, see Figure \ref{fig_det_spacer} (d).

\subsection{Design of the chamber and outer board}

\begin{figure}[!htb]
\begin{center}
\includegraphics[width=0.49\columnwidth]{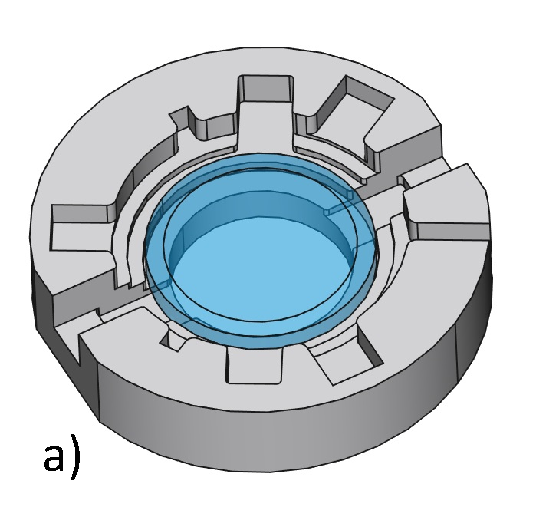}
\includegraphics[width=0.49\columnwidth]{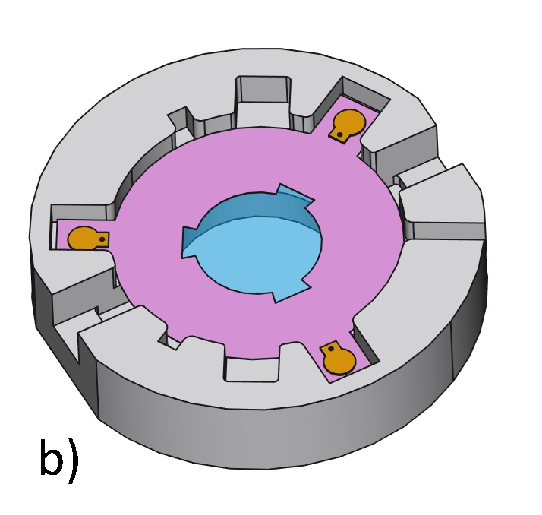}

\includegraphics[width=0.49\columnwidth]{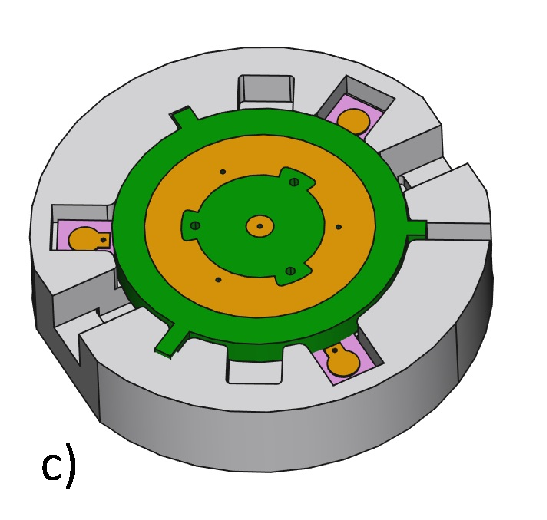}
\includegraphics[width=0.49\columnwidth]{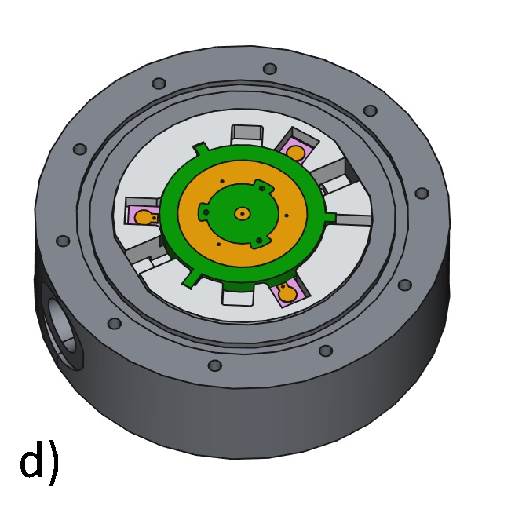}
\caption{ a)-c) Sketch of the detector parts assembly within the PEEK insert. MgF$_2$ crystal, spacer and Micromegas board are placed within the PEEK insert one after another without soldering. d) PEEK insert placed within the Al detector chamber}
\label{fig_det_assembly}
\end{center}
\end{figure}

The main structural component of the chamber is the cylindrical aluminium case that houses a non-conductive insert for placement of the MgF$_2$ crystal, spacer, and MM board, as shown in Figure \ref{fig_det_assembly}. The outer diameter of the case is 80~mm. The case also features the grooves for the o-ring seals on both the front and back sides. The insert was made from PEEK due to its good mechanical, electrical, and outgassing properties. A very important aspect of the design was the simplicity of the assembly/disassembly procedure to reduce the exposure time of sensitive parts to air, dust, and humidity. This is of particular importance when using CsI photocathodes, which are known for their fast degradation when exposed to air. The assembly is performed from the back side of the chamber by first placing a PEEK insert into the Al case. The PEEK insert is designed to have various indentations that ensure unique positioning of the spacer and MM board thus enabling precise mating of the electrical connections. Afterwards, the MgF$_2$ crystal, spacer, and MM board can be inserted one after another in their dedicated positions of the PEEK insert. The chamber is finally sealed from the back side with the OB, which is fastened to the Al case using ten M3 bolts.

\begin{figure}[!htb]
\begin{center}
\includegraphics[width=0.49\columnwidth]{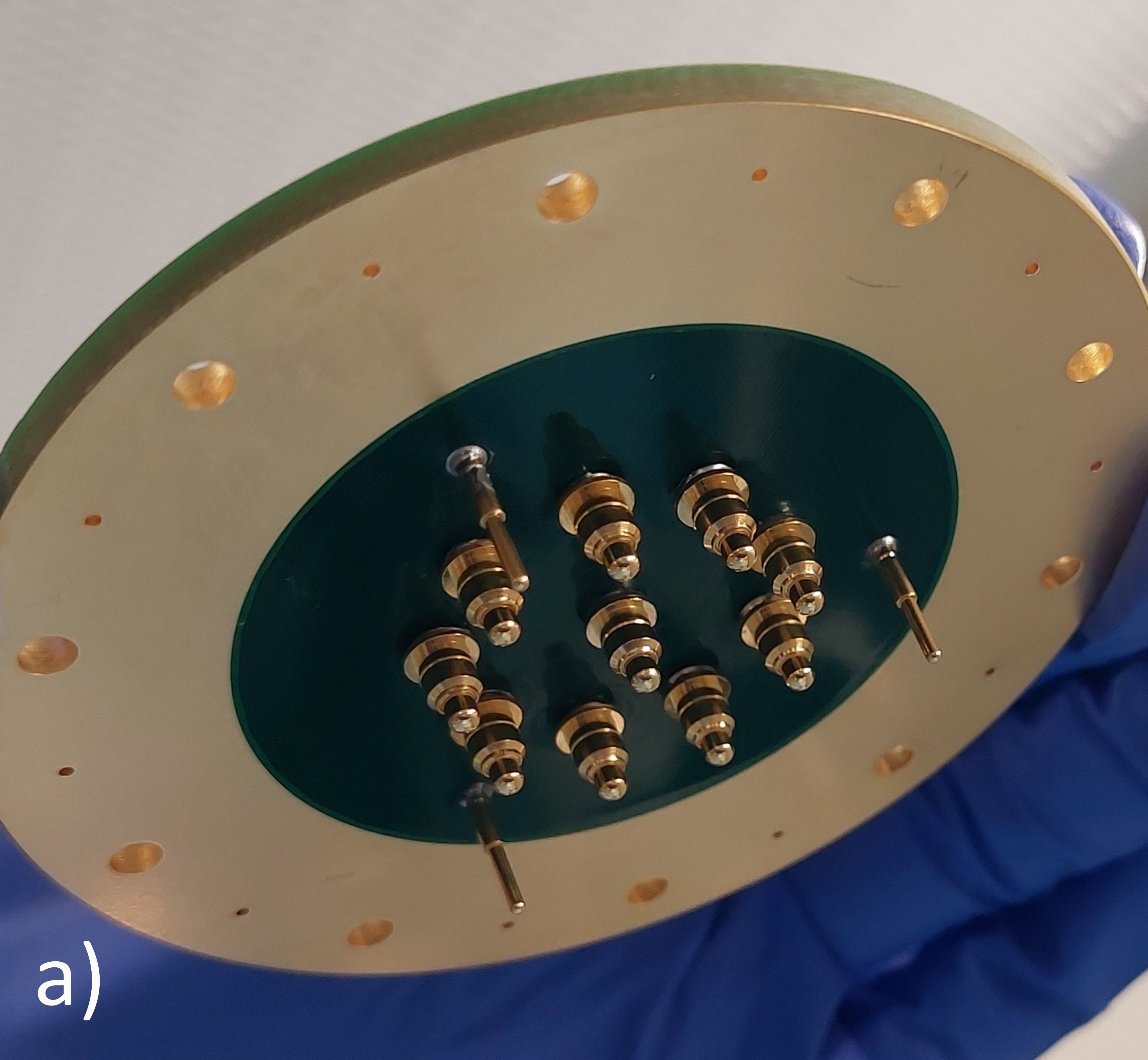}
\includegraphics[width=0.49\columnwidth]{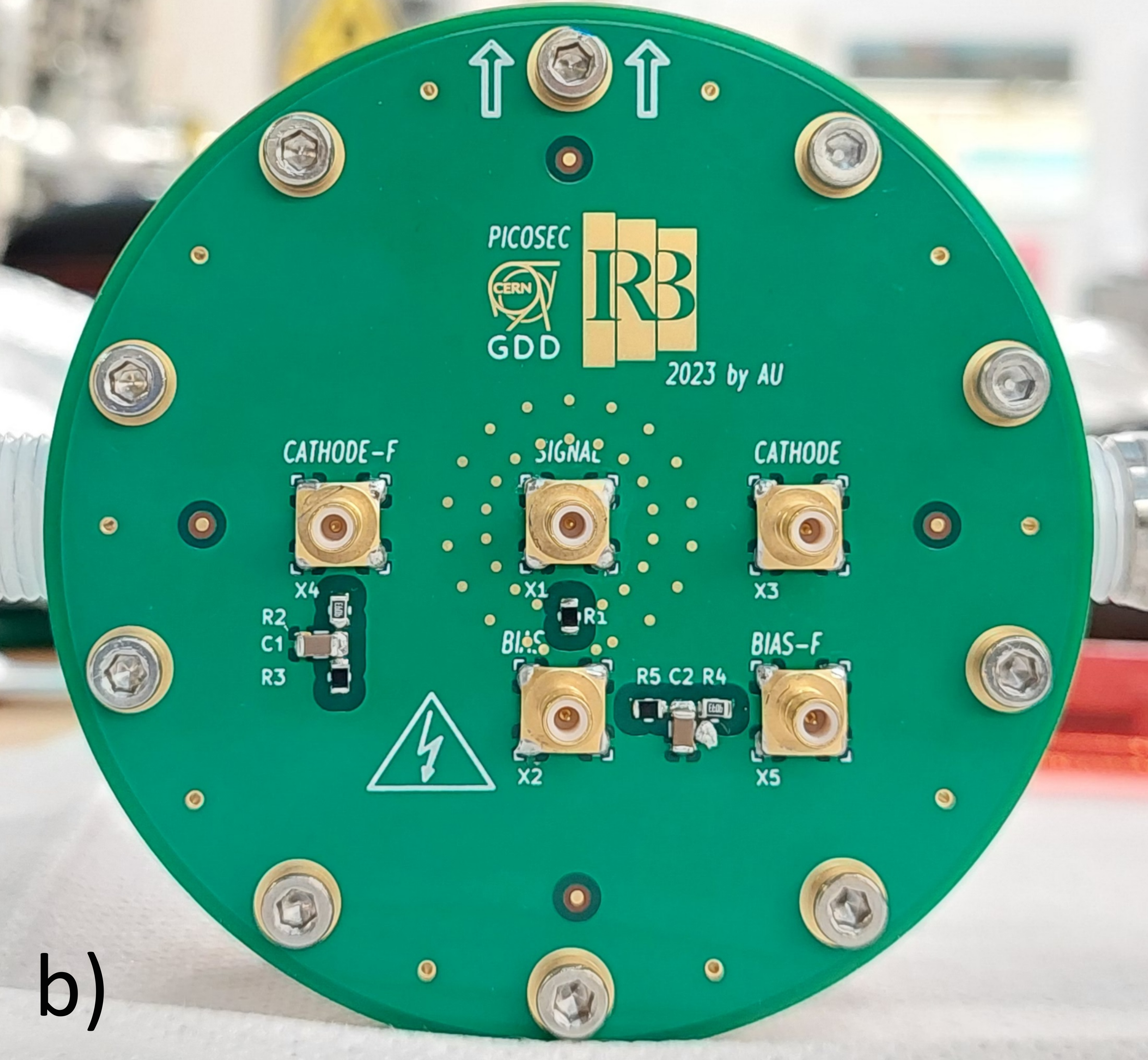}

\caption{ a) Outer board bottom side with soldered spring-loaded pins. b) Outer board top side with integrated HV filters and SMB connectors for cathode and anode connection to HV.}
\label{fig_OB}
\end{center}
\end{figure}

The OB is made as a 3.2~mm thick 4-layer FR4 PCB. The OB plays different roles that are important for achieving high performance in the detector operation: ensuring gas tightness, providing electrical connection to the outer world, and mechanically pressing the MM board against the crystal. Gas tightness is accomplished using the epoxy-filled blind and buried vias interleaved through the multiple layers, thus reducing the possible leakage spots. Moreover, when tightened to the housing, the OB compresses the 1.78~mm fluorocarbon o-ring gasket to seal the chamber. 
The electrical connection to MM board and photocathode is achieved using multiple spring-loaded pins mounted on the internal side of OB, see Figure \ref{fig_OB} (a). The outer side of the board contains the SMB connectors for signal and biasing and surface mount components for additional bias voltage filtering, see Figure \ref{fig_OB} (b). A lot of attention was paid to the layout features that influence the integrity of the signal and EM shielding of the detector internals. In this sense, the signal path was minimized by placing the signal readout SMB connector directly over the spring-loaded pin. Most of the copper within all layers was used as the ground plane to provide, both the shortest return path for the signal current and to minimize the noise pickup from external sources. In addition, the mating surface to the Al case is left conductive (gold plated) to accomplish a good electrical connection. Besides the electrical connection, usage of spring-loaded pins has shown to be an elegant and efficient way to achieve the adequate pressure of the MM board to the MgF$_2$ crystal. Constant and even pressure of the MM board to the crystal is required to obtain a well-defined drift gap thickness. For this purpose, nine extra heavy (153~g) spring-loaded pins placed in a circular pattern are used to evenly press the MM board against the crystal with a total pressing force of 1.4~kg. To minimize deformations of the board in the active area of the detector, a central spring-loaded pin was selected to have a pressing force of 23~g. Due to the larger distance of the inner side of the OB to the pads on the spacer bottom side, a third type of 3.5~mm longer spring-loaded pin is used for the photocathode electrical connection.  The total board thickness is 3.2 mm to reduce the mechanical deformations to an acceptable level during vacuum pumping.

\begin{figure}[!htb]
\begin{center}
\includegraphics[width=0.89\columnwidth]{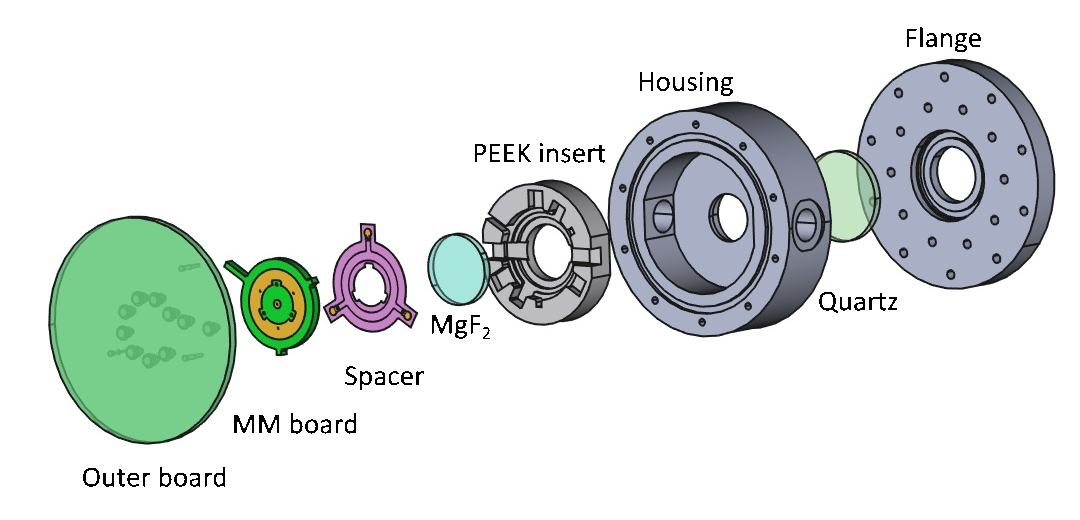}

\caption{ Exploded view from the detector's back side.}
\label{fig_exploded}
\end{center}
\end{figure}

A hollow flange with a 3~mm thick circular quartz window ($\varnothing$32 mm) seals the chamber from the front side and features the opening that allows the UV light to enter the detector volume for calibration measurements, see Figure \ref{fig_exploded}. 


\section{Experimental verification of the detector design}
\label{sec:chap4}
\subsection{Measurements of parasitic parameters of the detector}
From section \ref{sec:sig_int} the readout pad capacitance and signal path stray inductance were identified as the important parameters that strongly influence slew rate and, thus, the achievable time resolution. Although the lumped circuit from Figure \ref{fig_schema_L_C} is relatively simple, there is no procedure for directly measuring its parameters, namely $C_{pad}$, $L_{\sigma}$, and $C_{con}$ once the detector is assembled. Due to this limitation, the parameters of the lumped model were identified from the impedance measurement. This measurement was conducted in the frequency range from 10~MHz to 1.5~GHz using a Siglent SVA1032X VNA (Vector Network Analyzer) calibrated with a reference plane at the signal SMB connector on the outer board.

\begin{figure}[!htb]
 \footnotesize
\begin{center}
\includegraphics[width=0.99\columnwidth]{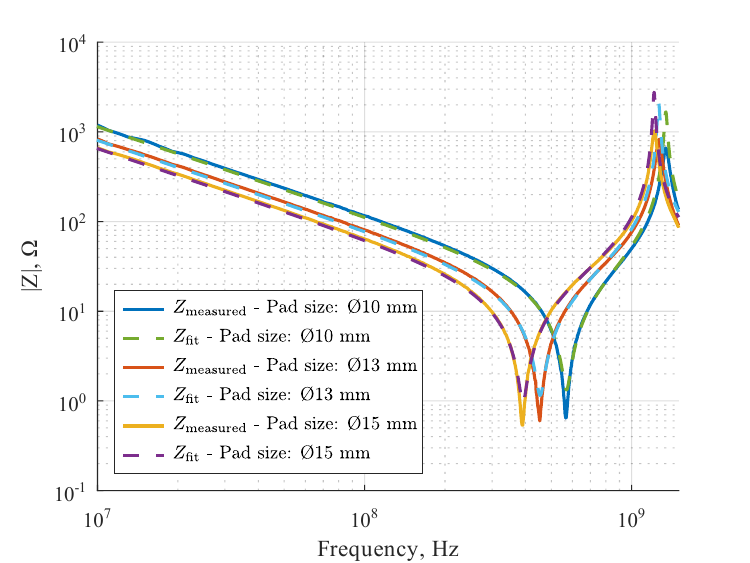}

\caption{Measured impedance of the detector with the fit overlayed.}
\label{fig_impedance_meas}
\end{center}
\end{figure}

The impedance of the detector seen from the output signal connector can be described with the following equation:

\begin{equation}
Z_{det}\left(j\omega\right)=\left(R_s+j\omega L_{\sigma}+\dfrac{1}{j\omega C_{pad}}\right)\left|\right|\dfrac{1}{j\omega C_{con}} \, ,
\label{eq_impedance_fit}
\end{equation}

where $\omega$ is the angular frequency and the parameter $R_s$ was added to model the relatively small resistance of the mesh and other conductive components in the signal path. The measured impedance characteristics for the detectors with a pad size of $\varnothing$10~mm, $\varnothing$13~mm, and $\varnothing$15~mm, together with the resulting fit to the $Z_{det}\left(j\omega\right)$ from equation \ref{eq_impedance_fit} are shown in Figure \ref{fig_impedance_meas}. Two resonances can be observed in the impedance characteristic. The first one, in the range between 350~MHz and 600~MHz is a series resonance between $C_{pad}$ and $L_{\sigma}$, and it greatly depends on the size of the pad. The second resonance is a parallel resonance between $C_{con}$ and $L_{\sigma}$ which does not change much between detector sizes, indicating their comparable stray inductance. All parameters obtained by the fitting procedure are listed in Table \ref{tbl_fit_params}. As expected, the pad capacitance varies with the pad area in the range between approx. 11~pF and 22~pF while the stray inductance is relatively constant around 7~nH.

\begin{table}[!htb]
\begin{center}
\caption{Impedance fit parameters}
\vspace{0.25 cm}

\label{tbl_fit_params}
\begin{tabular}{l|l|l|l}

\multirow{2}{*}{\begin{tabular}[c]{@{}l@{}}Fit \\ parameter\end{tabular}} & \multirow{2}{*}{\begin{tabular}[c]{@{}l@{}}Pad size:\\  $\varnothing$10 mm\end{tabular}} & \multirow{2}{*}{\begin{tabular}[c]{@{}l@{}}Pad size:\\  $\varnothing$13 mm\end{tabular}} & \multirow{2}{*}{\begin{tabular}[c]{@{}l@{}}Pad size:\\  $\varnothing$15 mm\end{tabular}} \\
            & & &\\ \hline \hline
$C_{pad}$, pF    & 11.43$\pm$0.27 & 17.24$\pm$0.41 & 21.91$\pm$0.52 \\ \hline
$L_{\sigma}$, nH & 6.81$\pm$0.17  & 7.12$\pm$0.17  & 7.54$\pm$0.18 \\ \hline
$C_{con}$, pF    & 2.52$\pm$0.08  & 2.57$\pm$0.07  & 2.55$\pm$0.07 \\ \hline
$R_s$, $\Omega$  & 1.33$\pm$0.30  & 1.15$\pm$0.26  & 0.96$\pm$0.23 \\ 
\end{tabular}
\end{center}
\end{table}

\subsection{Signal dynamics measurements}

The importance of minimizing the ratio between noise and slew rate (d$V$/d$t$) is clearly visible in equation \ref{eq_time_resolution}. Therefore, it was important to determine the relationship between the detector geometry and the dynamics of the output signal, especially the slew rate.

To exclude the effects of the photocathode's quantum efficiency, the measurements were performed under laboratory conditions using a single photoelectron and the same field configuration for all detector types. Single photoelectrons were generated with a Cr photocathode, and the UV light was introduced through the front side window of the detector. To reduce transmission line effects, a custom-built low-noise preamplifier board (650~MHz, 38~dB) was directly plugged into the signal port of the OB. The preamplifier is based on a SiGe HBT (Hybrid Bipolar Transistor), which was originally developed for diamond detectors \cite{hoarau2021rf}. The waveforms of the amplifier were recorded with the LeCroy WR8104 oscilloscope with a sampling rate of 10~GS/s, a bandwidth of 1~GHz, and a vertical scale of 20 mV/div. The thickness of the drift gap of 120$\pm$5~$\mu$m was the same for all measured detector prototypes.

\begin{figure}[!htb]
\begin{center}
\includegraphics[width=0.99\columnwidth]{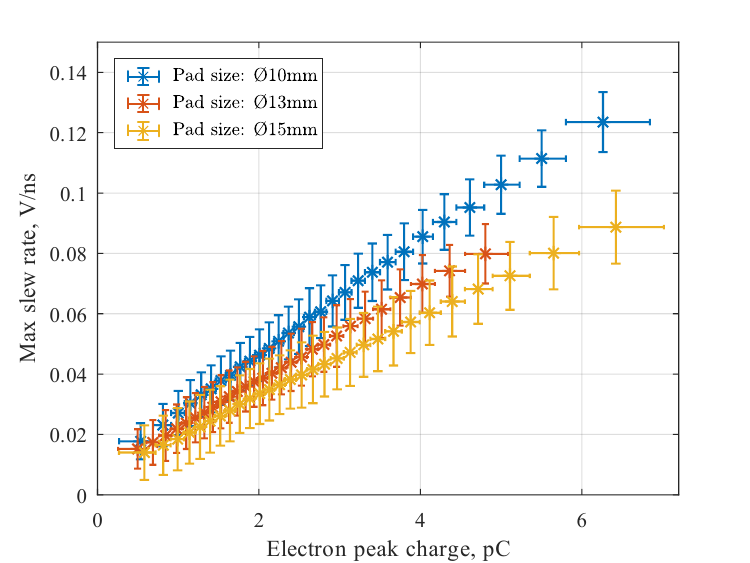}
\caption{ Maximum slew rate vs. electron peak charge for detectors with $\varnothing$10, $\varnothing$13, and $\varnothing$15 mm active area. The applied cathode and anode voltages were -475~V and 275~V, respectively. }
\label{fig_slwe_rate}
\end{center}
\end{figure}

The measured signals were processed and the maximum slew rate in the electron peak leading edge was selected as a measure for comparison. Since the slew rate depends on the signal amplitude, the extracted maximum slew rate was compared as a function of the electron peak charge. Figure \ref{fig_slwe_rate} shows the dependence of the maximum slew rate on the electron peak charge for the three measured detectors. As expected, the detector with the lowest pad capacitance has the highest maximum slew rate.

\begin{figure}[!htb]
\begin{center}
\includegraphics[width=0.99\columnwidth]{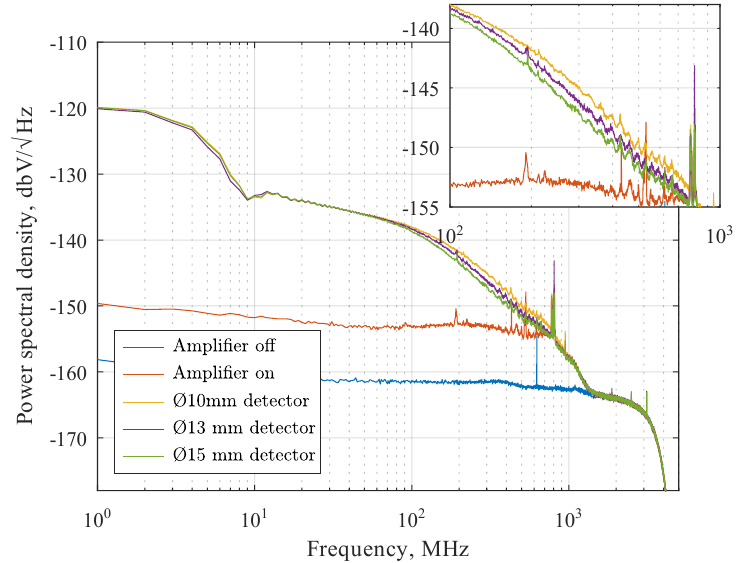}
\caption{Power spectral density of single photoelectron signals with the amplifier and digitizer noise floor.}
\label{fig_fft}
\end{center}
\end{figure}

Moreover, the power spectral density (PSD) was measured for single photoelectron events in the frequency range from 1~MHz to 1.5~GHz. Figure \ref{fig_fft} shows the PSD obtained by averaging 800 spectra of individual events together with the PSD of the amplifier and oscilloscope noise. Events with electron peak charges between 4~pC and 5~pC were selected to allow a comparison of detectors with different pad diameters across the entire frequency range. It can be observed that the largest difference between the spectra is in the frequency range from 100~MHz to 1~GHz, which corresponds to the fast electron peak part of the waveform. Both the slew rate and the frequency response show the advantages of using detectors with lower pad capacitance.

\subsection{Detector time response to 150~GeV muons}

\begin{figure}[!htb]
\begin{center}
\includegraphics[width=0.98\columnwidth]{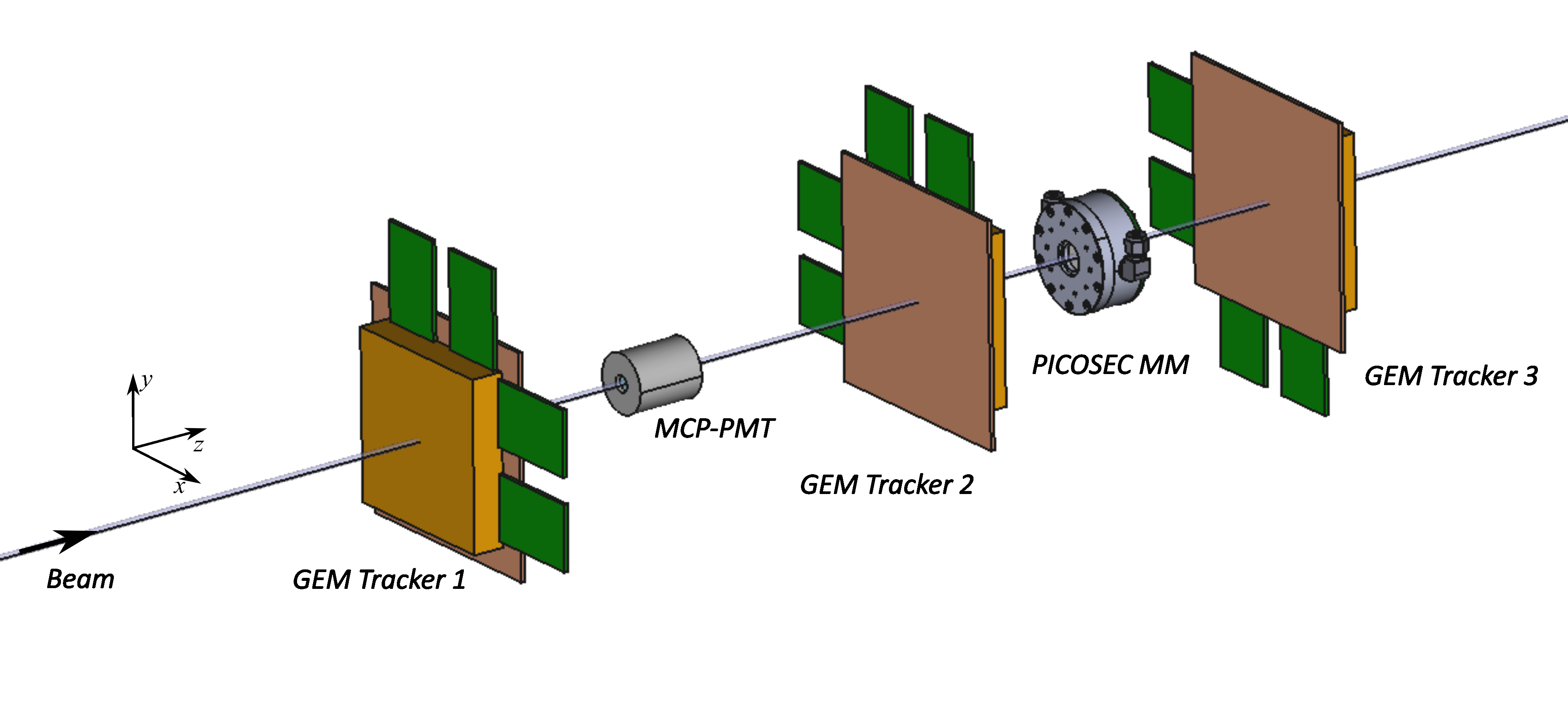}
\caption{ Schematic of the experimental setup for time response test of a single channel PICOSEC MM to 150~GeV/c muons.  }
\label{fig_beam_setup}
\end{center}
\end{figure}

The time response of the single-channel detectors for 150 GeV/c muons was measured at CERN at the secondary beamline H4 of the SPS during the RD51 and DRD1 test beam campaigns. As shown in Figure \ref{fig_beam_setup}, the experimental setup consisted of a tracker telescope based on three triple GEM detectors, used to determine the particle hit position, and an MCP-PMT\footnote{Hamamatsu Microchannel plate photomultiplier tube (MCP-PMT) R3809U-50 \url{https://www.hamamatsu.com/jp/en/product/type/R3809U-50/index.html}}, which served as a timing reference and trigger detector. All three single-channel PICOSEC MM prototypes were tested with an 18~nm thick CsI photocathode on a 2.38~nm thick conductive Cr substrate. The thickness of the drift gap was 120~$\pm$~5~$\mu$m, the same as in previous measurements with a single photoelectron.

The signal from the tested PICOSEC MM detector, the signal reference detector, and the serial bitstream with the event ID from the tracker were digitised and recorded on the same oscilloscope. The time stamps of PICOSEC MM and the reference detector were obtained by offline analysis. A fit to the signal's leading edge was performed with a generalised logistic function from which the timestamp was calculated at 20 \% constant fraction with the same methods as described in \cite{bortfeldt2018picosec, sohl2020development}. The time resolution of the detector was measured using the distribution of time differences between the time stamps of the PICOSEC MM and the reference detector. All time resolutions given in this paper include the time resolution of the reference detector.

To calculate the time difference distribution, several cuts were applied to the triggered events. A time window cut selects events within ~300 ps of the median time difference of all triggered events. Additional cuts are applied for minimum and maximum signal amplitude, and a geometric cut selects only events that occur within a certain diameter around the center of the pad.

\begin{table}[]
\begin{center}
\caption{Time resolution, electron peak charge, signal rise time, and noise of $\varnothing$10~mm, $\varnothing$13~mm, and $\varnothing$15~mm active area detector prototypes for cathode and anode voltage of -435~V and 265~V, respectively.}
\vspace{0.25cm}
\footnotesize
\resizebox{\textwidth}{!}{\begin{tabular}{l|ll|ll|ll}

\multirow{2}{*}{\begin{tabular}[c]{@{}l@{}}Detector\\ active area\end{tabular}} & \multicolumn{2}{l|}{\multirow{2}{*}{$\varnothing$10 mm}} & \multicolumn{2}{l|}{\multirow{2}{*}{$\varnothing$13 mm}} & \multicolumn{2}{l}{\multirow{2}{*}{$\varnothing$15 mm}} \\ 
& \multicolumn{2}{l|}{} & \multicolumn{2}{l|}{} &\multicolumn{2}{l}{}\\   \hline \hline

\multirow{2}{*}{Time resolution ($\sigma$), ps} & \multicolumn{1}{l|}{\begin{tabular}[c]{@{}l@{}}within\\ $\varnothing$4~mm \end{tabular}} & \begin{tabular}[c]{@{}l@{}}within\\ $\varnothing$9 mm\end{tabular} & \multicolumn{1}{l|}{\begin{tabular}[c]{@{}l@{}}within \\ $\varnothing$7~mm\end{tabular}} & \begin{tabular}[c]{@{}l@{}}within\\ $\varnothing$12~mm \end{tabular} & \multicolumn{1}{l|}{\begin{tabular}[c]{@{}l@{}} within \\ $\varnothing$9 mm \end{tabular}} & \begin{tabular}[c]{@{}l@{}}within\\$\varnothing$14~mm \end{tabular} \\ \cline{2-7} 
 & \multicolumn{1}{l|}{13.8$\pm$0.2} & 15.7$\pm$0.4                                                 & \multicolumn{1}{l|}{17.9$\pm$0.7}  & 20.0$\pm$0.3                                                   & \multicolumn{1}{l|}{17.8$\pm$0.9}  & 20.3$\pm$0.4  \\ \hline
e-peak charge, pC & \multicolumn{2}{l|}{19.2 $\pm$ 1.4} & \multicolumn{2}{l|}{15.3 $\pm$ 0.4} & \multicolumn{2}{l}{17.0 $\pm$ 0.6} \\ \hline
Rise time 10-90\%, ps & \multicolumn{2}{l|}{710 $\pm$ 5} & \multicolumn{2}{l|}{785 $\pm$ 3} & \multicolumn{2}{l}{822 $\pm$ 8} \\ \hline
Noise, mV & \multicolumn{2}{l|}{1.256 $\pm$ 0.010} & \multicolumn{2}{l|}{1.188 $\pm$ 0.004} & \multicolumn{2}{l}{1.232 $\pm$ 0.008} \\ \hline
\end{tabular}}
\end{center}

\label{tbl_time_res}
\end{table}

\begin{figure}[!htb]
\begin{center}
\includegraphics[width=0.49\columnwidth]{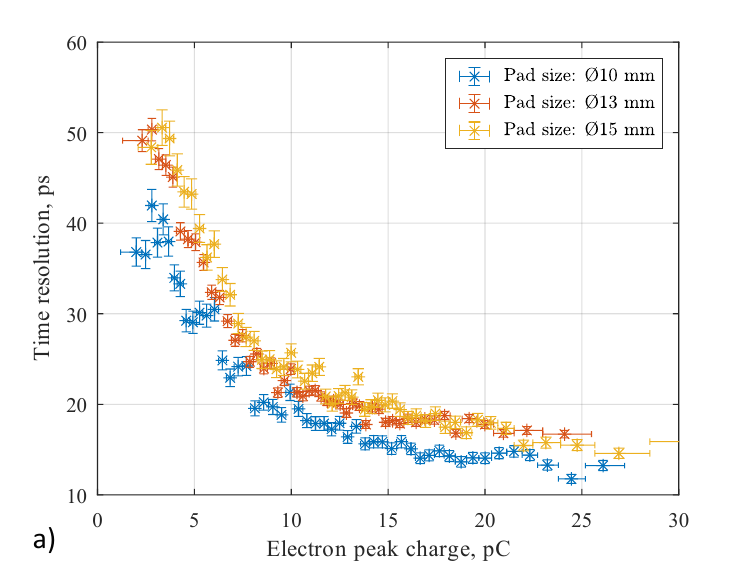}
\includegraphics[width=0.49\columnwidth]{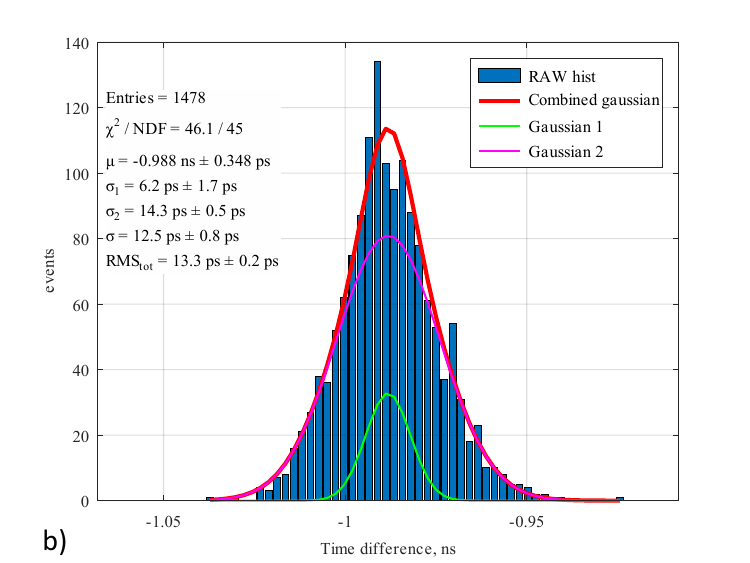}
\caption{a) Time resolution dependence on e-peak charge for $\varnothing$10~mm, $\varnothing$13~mm and $\varnothing$15~mm active area detector prototypes at cathode and anode voltage of -435~V and 265~V, respectively. b) Time difference distribution for MIPs and superimposed double Gauss fit within central $\varnothing$4~mm of the $\varnothing$10~mm active area detector in the best-performing run. Cathode and anode voltage was set to -415~V and 275~V, respectively.}
\label{fig_resolution}
\end{center}
\end{figure}

A double Gaussian fit (with equal mean values) is made to a time difference distribution of 150~GeV/c muons \cite{bortfeldt2018picosec}, and the time resolution is given as the standard deviation. For the 3~mm thick MgF$_2$ radiator, the projection of the Cherenkov light cone reaching a photocathode's surface is 6~mm in diameter. Signals from particles passing within the $\varnothing$4 mm, $\varnothing$7 mm, or $\varnothing$9 mm central pad region of the detector contain contributions from all generated photoelectrons. However, signals from tracks passing outside the central pad region have a lower amplitude due to the partial loss of photoelectrons outside the pad area. The time resolution was obtained within the central region and almost the entire active region (1~mm smaller in diameter than the pad size). To allow valid comparison, all three single-channel prototypes were evaluated with the same voltage settings, 265~V at the anode and -435~V at the cathode. The time resolution, noise, signal rise time, and electron peak charge are summarised in the Table \ref{tbl_time_res}. Figure \ref{fig_resolution} (a) shows the dependence of the time resolution on the electron peak charge for all three detectors, calculated from all detected events without the geometric cut.

It can be observed that the detector prototypes with a $\varnothing$13~mm and $\varnothing$15~mm active area showed similar time resolution, although it was expected that the detector with the smaller capacitance would outperform the one with the larger capacitance. However, due to the limitations during the beam tests, it was not possible to test all three prototypes with identical photocathode and under identical climatic conditions. From Table \ref{tbl_time_res} it can be seen that the $\varnothing$13~mm prototype had a lower gain, which influences the timing performance. On the other hand, the smallest detector with the lowest pad capacitance showed a significant improvement in time resolution for the same voltage settings, 13.8$\pm$0.2~ps for tracks passing within the $\varnothing$4~mm pad centre region and 15.7$\pm$0.4~ps over a $\varnothing$9~mm area. This can be attributed to the lowest capacitance leading to an improvement in signal dynamics and amplitude. Moreover, it was possible to further optimise the voltage setting on the 10~mm detector and achieve the excellent time resolution of 12.5~$\pm$~0.8~ps over a 4~mm area, see Figure \ref{fig_resolution} (b). It should be noted that the given time resolution also includes the time resolution of the reference detector which pushes the time resolution of the PICOSEC MM detector close to the 10~ps limit. It can be concluded that all measures taken during the design phase of the detector have led to an improvement in the timing performance of the PICOSEC MM detector technology.

It must be mentioned that the measured RMS noise of ~1.2~mV is dominated by the quantisation noise of the 8-bit ADC in the oscilloscope in the range of 50~mV/div, which is required to cover the dynamic range. Given that the RMS noise of the amplifier at 20~mV/div (Figure \ref{fig_fft}) is 0.827$\pm$0.043~mV, there is still room for improving the SNR and thus the time resolution if higher resolution digitisers (e.g. 12-bit) are used.

\section{Conclusion}
\label{sec:chap7}
This paper provides guidelines for the design of a single-channel PICOSEC MM detector assembly that can be used as a test platform for future detector optimization studies. It covers the design of the detector board, vessel, auxiliary mechanical parts and electrical connections for high voltage and signals with the aim of improving stability, reducing noise, and ensuring signal integrity to maximise timing performance. In addition, the proposed design is characterised by a simple and fast assembly procedure that allows for quick replacement of the detector internals.

The article also describes the influence of pad capacitance and stray inductance on the signal dynamics and emphasises the importance of controlling both capacitance and inductance in the design phase.

To validate the design procedure, a prototype detector assembly and three interchangeable PICOSEC MM detector boards with different readout pad diameters were manufactured. As part of the initial tests, a method was developed to measure pad capacitance and stray inductance based on an impedance fit over a wide frequency range. In addition, the signal dynamics were verified for all three detectors by the single photoelectron test in the laboratory environment in both the time and frequency  domain, which qualitatively confirmed the predictions.

Finally, the timing performance of the detectors with different pad sizes was verified in the MIP beam test. All developed detectors showed stable operation under MIP irradiation, and excellent timing performance with time resolution well below 20~ps was observed. Most importantly, with the smallest detector of $\varnothing$10~mm, a record resolution of 12.5~ps was achieved for the PICOSEC technology, justifying the meticulous design of all detector components.

The same housing with the outer board, spring-loaded pins and minimised mechanical coupling of detector board and photocathode has already been successfully tested with resistive PICOSEC MM and $\mu$RWELL detector topologies.

\section*{Acknowledgements}
We thank the CERN-EP-DT-MPT Workshop, particularly Antonio Teixeira, Bertrand Mehl and Rui de Oliveira for the useful discussion and production of MM prototypes. We acknowledge the financial support of the EP R$\&$D, CERN Strategic Programme on Technologies for Future Experiments; the RD51 Collaboration, in the framework of RD51 common projects; the Cross-Disciplinary Program on Instrumentation and Detection of CEA, the French Alternative Energies and Atomic Energy Commission; the PHENIICS Doctoral School Program of Université Paris-Saclay, France; the Program of National Natural Science Foundation of China (grant number 11935014, 12125505); the COFUND-FP-CERN-2014 program (grant number 665779); the Fundação para a Ciência e a Tecnologia (FCT), Portugal; the Enhanced Eurotalents program (PCOFUND-GA-2013-600382); the US CMS program under DOE contract No. DE-AC02-07CH11359.

\vspace{0.5 cm}
\bibliographystyle{elsarticle-num} 
\bibliography{literatura}

\end{document}